\documentstyle[twocolumn,prb,aps,epsf]{revtex}
\tighten
\begin{document}
\title{
Luminescence from highly excited nanorings:
Luttinger liquid description} 
\author{T. V. Shahbazyan}
\address{Department of Physics and Astronomy,
Vanderbilt University, Nashville, TN 37235}
\author{I. E. Perakis$^{*}$}
\address{Department of Physics, University of Crete,
P.O. Box 2208, 710 03, Heraklion, Crete, Greece}
\author{M. E. Raikh}
\address{Department of Physics, University of Utah, Salt Lake City,
UT 84112}
\maketitle
\draft
\begin{abstract}
We study theoretically the luminescence from quantum dots of a ring geometry.
For high excitation intensities, photoexcited electrons and holes
form Fermi seas. Close to the  emission threshold, the 
single-particle spectral lines aquire weak many-body satellites.
However, away from the threshold, the discrete luminescence spectrum
is completely dominated by many-body transitions.
We employ the Luttinger liquid approach to {\em exactly} calculate the 
intensities of {\em all} many-body spectral lines. We find that the
transition from single-particle to many-body structure of the emission
spectrum is governed by a single parameter and that the distribution of
peaks away from the threshold is {\em universal}.
\end{abstract}

\section{Introduction}

Photoluminescence (PL) from zero-dimensional objects (quantum dots)
is one of the highlights in physics of nanostructures which emerged during the
last decade.  Early papers 
(see, e.g., Refs.\ \onlinecite{brunner92,marzin94}, and the review 
article Ref. \onlinecite{zrenner00}) reported the PL spectra consisting 
of "zero-width" luminescence lines. High surface density of quantum dots
caused an ambiguity in assigning of these lines.
In the later studies the emission from a {\em single} dot was resolved.  
This progress\cite{gammon00} has permitted the PL spectroscopy of individual
dots with controllable exciton  population determined  by
the excitation intensity, 
and also at tunable charge states.\cite{warburton00,regelman01}

At low excitation intensity the number of excitons in a dot is
either one or zero. Then the emission line corresponds to the transition
between the lowest size-quantization levels in the conduction and the valence
bands. Upon increasing the excitation intensity, the number of excitons in
a dot, $N$, can be larger than one. This leads to the features in 
the PL spectrum which  must be interpreted in terms of
recombination within many-exciton complexes.\cite{ikezawa97,landin98,bayer98,kulakovskii99,dekel98,dekel00a,dekel00b,dekel01,findeis00,bayer00}
Different recombination processes within a complex
result in a multitude of the emission lines. 
One reason for an emergence of additional PL lines is that,
for $N>2$, the carriers constituting the complex occupy higher 
size-quantization levels. Another reason, is that the 
the interactions between the strongly confined 
photoexcited carriers lift the {\em degeneracies} of the final
many-body states. The latter mechanism of interaction-induced
multiplication of the emission lines was addressed in 
Refs.\ \onlinecite{dekel98,dekel00a,hawrylak99}
for the situations with\cite{hawrylak99}  and without\cite{dekel98,dekel00a} 
orbital degeneracy of single-particle states. 
The calculations carried out  in 
Refs.\ \onlinecite{dekel98,dekel00a,hawrylak99} predicted
the splittings  of the PL
lines, originating from different many-body final states,
to be of the order of the matrix element of the interaction 
potential. The actual positions of the lines predicted 
by these calculations reproduce quite accurately the 
experimental PL spectra of 
Refs.\ \onlinecite{dekel98,dekel00a,dekel00b,dekel01} (for up to $N=16$) 
and of Ref.\onlinecite{bayer00} (for $1\leq N \leq 6$).

Within the approaches of 
Refs.\ \onlinecite{dekel98,dekel00a,hawrylak99},
each many-body line corresponds to recombination of an electron
and a hole having  the {\em same} size-quantization quantum numbers.   
We note here that in the experiment\cite{dekel98,dekel00a}
additional emission lines have been observed
that were identified with the
transitions between {\em different} size-quantization
levels. These transitions originate from a
{\em shake-up} effect in a confined electron-hole system.
Namely,  the radiative recombination of an
electron-hole pair is accompanied by an  {\em internal} excitation
within the exciton multiplex. These many-body processes
are missed within the theoretical 
approaches of Refs.\ \onlinecite{dekel98,dekel00a,hawrylak99}. 
Meanwhile, as we argue below, the PL lines originating from
the shake-up processes multiply very rapidly with increasing
$N$, and for high enough $N$ become the dominant feature of the
PL spectrum. 
Here we develop a theory of the many-body luminescence
 from a quantum dot for the limiting case $N\gg 1$.

When a number of photoexcited carriers is large, an adequate 
description of PL from a dot must be developed in terms of 
Fermi seas formed by  equal numbers (determined by excitation)
of electrons and holes in  conduction  and valence bands,
respectively (see inset in Fig.\ \ref{fig:1}).
As in the case of a small number of carriers, such a description 
is based on the fact that PL is preceeded by a fast nonradiative
relaxation of electrons and holes into the corresponding ground 
states.\cite{dekel98,dekel00a,dekel00b,dekel01,findeis00,bayer00}
Microscopic mechanism of this relaxation is addressed, e.g., in 
Ref.\ \onlinecite{Toda}.

For noninteracting system, the emission lines would correspond to
transitions between size-quantization levels in conduction and valence
bands which obey the selection rules. Within this description, the
single-particle emission spectrum near the Fermi edge is given by the
Golden rule,
\begin{eqnarray}
\label{non-inter}
I(\omega)\propto 
\sum_n C_n\delta \Bigl[\omega+(\Delta_1+\Delta_2) n\Bigr],
\end{eqnarray}
where $\Delta_1$ and $\Delta_2$ are level spacings for electrons
and holes; $C_n$ are the oscillator strength which depend on $n$
only {\em weakly} ($\omega<0$ is measured from the Fermi edge). 

As discussed above, the many-body transitions, resulting from interactions of
carriers in a dot, change {\em qualitatively} the form of the spectrum. 
Analogously to recombination within an exciton multiplex,
here a removal of an e-h pair shakes up the respective Fermi seas by
causing them to emit Fermi sea excitations.
Since in a {\em finite} system, the energies of 
excitations are {\em quantized}, such a shake-up
would lead to the spectrum of a form
\begin{equation}
\label{modified}
I(\omega)\propto\sum_{mn} C_{mn}
\delta \Bigl(\omega+m\tilde{\Delta}_1+n\tilde{\Delta}_2 \Bigr),
\end{equation}
rather than Eq.\ (\ref{non-inter}). Here $\tilde{\Delta}_1$ and
$\tilde{\Delta}_2$ are the level spacings 
renormalized by interactions. All the information about many-body 
correlations in the system is encoded in the oscillator strengths
$C_{mn}$. As we will see below, $C_{mn}$, being governed by
interactions,  are {\em strong functions} of $m$ and $n$.

The goal of the present paper is to demonstrate that the
oscillator strengths $C_{mn}$ can be evaluated {\em analytically}
for a dot of a ring geometry. Such dots have been recently 
fabricated\cite{warburton00} and their emission spectra (including
many-body effects) were studied for low excitation intensities both
experimentally\cite{lorke00,peterson00} and 
theoretically.\cite{chaplik95,rudo00,HuiHu00,HuiHu01a,HuiHu01b,ulloa00}
Our approach is valid when the number of carriers, N, is large enough 
(the accuracy is 1/N).  However, this "asymptotic" consideration 
allows us to establish the {\em universal} properties of the many-body
spectrum away from the threshold.

For a ring-shaped dot, the electron and hole Fermi seas represent
one-dimensional (1D) systems. This allows us to use the finite-size
Luttinger-liquid description\cite{schulz95} for calculation of the
emission spectrum.
Note that the Luttinger liquid model was  employed earlier for
calculations of the Fermi-edge optical properties of {\em infinite}
1D systems (with and without defects) in 
Refs. \onlinecite{gogolin93,prokof'ev94,kane94,sassetti98,kramer00}.

We show that due to a finite size of the system, the structure of the emission
spectrum is different in low- and high-$\omega$ domains. Namely, for 
$\mu\ln\bigl|\omega/(\tilde\Delta_1+\tilde\Delta_1)\bigr|\ll 1$
(low frequencies) the spectrum is dominated by single-particle peaks;
many-body satellites have a relative magnitude $\sim \mu$, where
$\mu\ll 1$ is the dimensionless interaction strength (Luttinger liquid
parameter). For high frequencies (i.e., 
$\mu\ln\bigl|\omega/(\tilde\Delta_1+\tilde\Delta_1)\bigr|\gg 1$),
the many-body peaks {\em completely dominate} the spectrum; roughly
speaking, in the high-$\omega$ domain, the oscillator strengths of 
single-particle peaks are evenly
distributed  among the multitude of many-body peaks.
Furthermore, the peaks cluster into groups
(generations), so that the patterns of peaks within the neighboring 
generations are almost identical.

\section{Emission from a Luttinger liquid ring}

We start with the two-component Luttinger liquid model on a 
ring\cite{shahbazyan97} with  
Hamiltonian $H_1+H_2+H_{int}$, where $H_j$ describe noninteracting
electrons ($j=1$) and holes ($j=2$) with linearized dispersions
(the slopes are determined  by the Fermi velocities $v_j$); $H_{int}$ 
describes the interactions between carriers via screened potential $U(x)$. 
The {\em e-h} recombination rate is given by
the  Golden rule
\begin{eqnarray}
\label{golden}
W(\omega)=
&&
\frac{2\pi}{L}\sum_f|\langle f |T|i\rangle|^2
\delta(E_i-E_f-\omega)
\nonumber\\
&&
=\frac{1}{L}\int_{-\infty}^{\infty}dt
e^{-i\omega t}\langle i|T^{\dag}(t)T(0)|i\rangle,
\end{eqnarray}
where $E_i$ and $E_f$ are the energies of initial (ground) 
and final (with {\em e-h} pair removed) states, and 
\begin{eqnarray}
\label{tran}
T=T_{+}+T_{-}
\mbox{\hspace{5mm}}
T_{\pm}=d\int_{0}^{L}dx \psi_{2\mp}(x)\psi_{1\pm}(x)
\end{eqnarray}
is the dipole transition operator. Here $\psi_{i\pm}$ are
annihilation operators for left 
($-$) and right ($+$) moving carriers,
$d$ is the interband dipole matrix element, and $L$ is the
ring circumference. Note that recombination occurs between left
(right) electrons  and right (left) holes. 
The recombination rate is then expressed via a four-particle Green
function, 
\begin{eqnarray}
\label{prob-D}
W(\omega)=
&&
d^2\int_{0}^{L}dx\int_{-\infty}^{\infty}dt
e^{-i\omega t}\bigl[ D_{+}(x,t)+D_{-}(x,t)\bigr]
\nonumber\\
=
&&
d^2\bigl[ D_{+}(\omega)+D_{-}(\omega)\bigr],
\nonumber\\
D_{\pm}(x,t)=
&&
\langle
\psi_{2\mp}^{\dag}(x,t)\psi_{1\pm}^{\dag}(x,t)
\psi_{1\pm}(0)\psi_{2\mp}(0)\rangle.
\end{eqnarray}
In order to evaluate $D_{\pm}(x,t)$ for a two-component Luttinger
liquid,\cite{larkin74,matveev93,penc93} we use the bosonization
technique on a ring\cite{shahbazyan97} (see Appendix A).
The final result reads
\begin{eqnarray}
\label{D-bos}
D_{\pm}(x,t)=
&&
\frac{e^{-it\delta_P}}{L^2}\,
\epsilon^{2(\mu_2+\mu_2)}
\nonumber\\ && \times
\Bigl[f_{\pm}(z_{1\pm})\Bigr]^{1+\mu_1}
\Bigl[f_{\mp}(z_{1\mp})\Bigr]^{\mu_1}
\nonumber\\ && \times
\Bigl[f_{\mp}(z_{2\mp})\Bigr]^{1+\mu_2}
\Bigl[f_{\pm}(z_{2\pm})\Bigr]^{\mu_2},
\end{eqnarray}
where  $\delta_P=\pi (v_1+v_2)/L$ 
is the energy shift (to be absorbed into the frequency) due the change in the
parity of electron and hole numbers, and $\epsilon$ is a cutoff. The coordinate
dependence of $D_{\alpha}(x,t)$ is determined by ($\alpha=\pm$)
\begin{eqnarray}
\label{I-bos}
f_{\alpha}(z_{j\alpha})
=
\frac{1}{1-e^{i\alpha (2\pi z_{j\alpha}/L +\alpha i\epsilon)}},
~~~
z_{j\alpha}=x-\alpha\tilde{v}_jt.
\end{eqnarray}
The explicit expressions for renormalized Fermi velocities $\tilde{v}_j$ 
and interaction-induced exponents  $\mu_j$ are given in Appendix \ref{app:a}.
Correspondingly, the level spacings are now 
$\tilde{\Delta}_j=2\pi \tilde{v}_j/L$.
The interaction strength is characterized by the ratio
$u/v_j$, with
\begin{eqnarray}
\label{u}
u= \frac{1}{\hbar \pi} \int dx U(x)
\end{eqnarray}
being the Fourier of screened potential; this ratio represents the average
(screened) interaction in units of the (bare) level spacing near the Fermi
energy. For weak interactions,
$u/v_j\ll 1$, we have (see Appendix \ref{app:a})
\begin{eqnarray}
\label{weak}
\mu_j
\simeq (u/4 v_j)^2,
\mbox{\hspace{5mm}}
\tilde{\Delta}_j\simeq \Delta_j(1+u/2v_j).
\end{eqnarray}
The  correlator $D_{\alpha}(x,t)$ is periodic in
variables $z_{j\alpha}$. In order to 
carry out the integration in Eq.\ (\ref{prob-D}) 
we first perform the Fourier expansion of 
$\bigl[f_{\alpha}(z_{j\alpha})\bigr]^{\nu}$ as
\begin{equation}
\label{expansion}
\Bigl[f_{\pm}(z_{j\pm})\Bigr]^{\nu}=
\frac{\sin \pi\nu}{\pi}
\sum_n B(n+{\tiny \nu},1-{\tiny \nu})
\,e^{\pm 2\pi i nz_{j\pm}/L},
\end{equation}
where $B(x,y)$ is the Beta-function. Substituting this expansion
into Eq.\ (\ref{D-bos}), and then 
into Eq.\ (\ref{prob-D}), we 
arrive at the expected form Eq.\ (\ref{modified}) of the
emission spectrum  
with the coefficients $C_{mn}$ cast in the following  closed
form (see appendix \ref{app:b}).
 \begin{eqnarray}
 \label{Cmn-int}
 C_{mn}=
 \int_{-\pi}^{\pi}
 &&
 \frac{d\phi_1d\phi_2d\phi_3}{(2\pi)^3}
 \frac{\epsilon^{2(\mu_1+\mu_2)}\,
 e^{-\frac{i}{2}(\phi_1+\phi_2)(m+n)}}
 {\Bigl(1-e^{i\phi_1}\Bigr)^{1+\mu_1}
 \Bigl(1-e^{i\phi_2}\Bigr)^{1+\mu_2}}
 \nonumber\\ && \times
 \frac{e^{-\frac{i}{2}\phi_3(m-n)}}
 {\Bigl(1-e^{i(\phi_2-\phi_3)}\Bigr)^{\mu_1}
 \Bigl(1-e^{i(\phi_1+\phi_3)}\Bigr)^{\mu_2}}.
 \end{eqnarray}
Note, that the sum in Eq.\ (\ref{modified}) is constrained by the
selection rule that $m$ and $n$ are of the same parity, i.e., the
combinations 
\begin{equation}
N=(m+n)/2,
~~~
M=(m-n)/2,
\end{equation}
which enter into the rhs of Eq.\ (\ref{Cmn-int}), are integers.
This is the result of the linear dispersion of
electrons and holes near the Fermi levels.

Formula (\ref{Cmn-int}) for the oscillator strengths is the main
result of this section. It is easy to see that it correctly reproduces
the non-interacting limit. Indeed, upon setting $\mu_i=0$, the 
integral over $\phi_3$ yields $C_{mn}=\delta_{mn}$.
Another important limiting case $m,n\gg 1$  corresponds to the 
transitions well away from the Fermi edge. In this case, the main
contribution to the integral (\ref{Cmn-int}) comes from the domain  
$\phi_1+\phi_2\sim (m+n)^{-1}\ll 1$. Within this domain, one can  
neglect the difference between $\phi_1$ and $-\phi_2$ in the last two
factors in the denominator. Then the integrals over $\phi_1,\phi_2$ 
factorize, yielding
\begin{eqnarray}
\label{Cmn-est}
C_{mn}=
\frac{\Gamma(N+1+\mu_1)\Gamma(N+1+\mu_2)}
{\Gamma(1+\mu_1)\Gamma(1+\mu_2)[\Gamma(N+1)]^2}\, K(M),
\end{eqnarray}
with 
\begin{eqnarray}
\label{Kn}
K(M)=
&&
\int_{-\pi}^{\pi}\frac{d\phi}{2\pi}
\frac{\epsilon^{2(\mu_1+\mu_2)}\, e^{iM\phi}}
{\Bigl(1-e^{-i\phi}\Bigr)^{\mu_1}
\Bigl(1-e^{i\phi}\Bigr)^{\mu_2}}
\nonumber\\ &&
=\frac{\epsilon^{2(\mu_1+\mu_2)}(-1)^M\Gamma(1-\mu_1-\mu_2)}
{\Gamma(1-M-\mu_1)\Gamma(1+M-\mu_2)},
\end{eqnarray}
where $\Gamma(x)$ is the Gamma-function.
It can be seen from Eq.\ (\ref{Kn}) that, for a given $N$, the
oscillator strengths, $C_{mn}$, fall off as  
$C_{mn}\propto |M|^{\mu_1+\mu_2-1}$ with increasing 
$|M|=\frac{1}{2}\vert m-n \vert$.
This slow power-law decay reveals strong correlations within
electron-hole system on a ring. Finally, using the large $x$
asymptotics of $\Gamma(x)$, we obtain the expression for the 
oscillator strengths valid for $N,|M|\gg 1$,
\begin{equation}
\label{asym}
C_{mn}=\frac{\epsilon^{2\mu}\Gamma(1-\mu)}
{\Gamma(1+\mu_1)\Gamma(1+\mu_2)}\frac{\sin\pi\tilde\mu}{\pi}
N^{\mu}|M|^{\mu-1},
\end{equation}
where $\mu=\mu_1+\mu_2$, and $\tilde \mu =\frac{1}{2}\mu+
\frac{1}{2}(\mu_1-\mu_2){\rm sgn} M$.

\section{Many-body structure of the emission spectrum}

The general expression (\ref{Cmn-int}) determines the heights of the
emission peaks, while the {\em order} of the peaks with 
different $\{m,n\}$ is governed by the $\delta$-functions in
Eq.\ (\ref{modified}), which ensure the energy conservation. 
Therefore, this order depends crucially on the relation
between $\tilde{\Delta}_1$ and $\tilde{\Delta}_2$. Moreover, 
a {\em commensurability} between $\tilde{\Delta}_1$ and 
$\tilde{\Delta}_2$ leads to accidental degeneracies in 
the positions of the emission lines. 
However, in order to establish the general properties
of the spectrum, it is instructive to consider first two 
particular cases of commensurate $\tilde{\Delta}_1$ and 
$\tilde{\Delta}_2$.    

We start with the symmetric case 
$\tilde{\Delta}_i=\tilde{\Delta}/2$ (and, hence, $\mu_i=\mu/2$). 
The peak positions, as determined by Eq.\ (\ref{modified}), coincide
with those for single-particle transitions, 
$|\omega|=N\tilde{\Delta}$. The corresponding oscillator strengths 
can be straightforwardly evaluated from Eq.\ (\ref{Cmn-int}) as
\begin{eqnarray}
\label{cN-exact}
c_N=\sum_{M}C_{N+M,N-M}=\Biggl[\int_{-\pi}^{\pi}\frac{d\phi}{2\pi}
\frac{\epsilon^{\mu}\, e^{-iN\phi}}
{\bigl(1-e^{i\phi}\bigr)^{1+2\mu}}\Biggr]^2.
\end{eqnarray}
For $N\gg 1$, the denominator of the integrand  can be expanded, yielding
\begin{eqnarray}
\label{cN}
c_N
\simeq (\epsilon N)^{2\mu}=
\biggl|\frac{\epsilon\omega}{\tilde{\Delta}}\biggr|^{2\mu}.
\end{eqnarray}
Note that single-particle oscillator strengths correspond to $c_N=1$.
We thus conclude that interactions affect strongly the peak heights
for $|\omega/\tilde{\Delta}|^{2\mu}\gg 1$, i.e., in the high frequency
domain. 
In fact, even for an arbitrary relation between $\tilde\Delta_1$ and
$\tilde\Delta_2$, the crossover between ``single-particle'' and
``many-body'' domains of the spectrum is governed by the
dimensionless parameter  
$\mu\ln \bigl|\omega/(\tilde{\Delta_1}+\tilde{\Delta_2})\bigr|$.

Indeed, consider now the case $\tilde\Delta_1=3\tilde\Delta_2$ 
(and thus $\mu_2\simeq 9\mu_1$) which renders a spectrum richer than
(\ref{non-inter}) and (\ref{cN}). As follows from 
Eq.\ (\ref{modified}), the spectral positions of the peaks are given
by $|\omega|/(\tilde{\Delta}_1+\tilde{\Delta}_2)=l/2$, where $l$ is an
integer. The  corresponding oscillator strengths can be evaluated
explicitly from Eq.\ (\ref{Cmn-int}) (see appendix \ref{app:c}).
The final result reads
\begin{eqnarray}
\label{D=2}
D_{\pm}(\omega)
=\frac{2\pi}{L}
&&
\biggl|\frac{\epsilon\,\omega}{\tilde{\Delta}}\biggr|^{2\mu}
\sum_l\Biggl[
\frac{1+\bigl|\frac{2\omega}{\tilde{\Delta}}\bigr|^{-\mu}}{2}
\,\delta(\omega+\tilde{\Delta} l)
\nonumber\\ &&
+
\frac{1-\bigl|\frac{2\omega}{\tilde{\Delta}}\bigr|^{-\mu}}{2}
\,\delta\Bigl[\omega+\tilde{\Delta} (l+1/2)\Bigr]
\Biggr],
\end{eqnarray}
with $\mu=\mu_1+\mu_2\simeq 10\mu_1$ and
$\tilde{\Delta}=\tilde{\Delta}_1+\tilde{\Delta}_2=4\tilde\Delta_1$.

The above result illustrates how the structure of the spectrum evolves
as the frequency departs from the Fermi edge.
For $\mu\ln\bigl|\frac{\omega}{\tilde{\Delta}}\bigr|\ll 1$, each 
single-particle peak, $|\omega|=l\tilde\Delta$
acquires a weak many-body satellite at $|\omega|=(l+\frac{1}{2})\tilde\Delta$.
In the opposite limit, $\mu\ln\bigl|\frac{\omega}{\tilde{\Delta}}\bigr|\gg 1$,
the oscillator strength of an ``integer'' peak gets equally 
redistributed
between the components of the doublet. The crossover frequency, 
$\Omega$, separating the ``single-particle'' and the developed many-body
domains of the spectrum is determined by the condition 
$\ln\bigl|\frac{\Omega}{\tilde{\Delta}}\bigr|\sim \mu^{-1}$.
The spectrum (\ref{D=2})
is schematically depicted in Fig.\ \ref{fig:1}. 

Let us turn to the 
structure of the spectrum in the general case of incommensurate 
$\tilde \Delta_1$ and $\tilde \Delta_2$.
We start from the observation that the peak positions can be 
classified by ``generations''. 
Namely, once a peak $\{m,0\}$ (or $\{0,n\}$)
emerges at $\omega=\omega_m=-m\tilde\Delta_1$ 
(or $\omega=\omega_n=-n\tilde\Delta_2$), it is followed by
next generations of peaks $\omega_m^{(k)}=\omega_m-k(\tilde\Delta_1+
\tilde\Delta_2)$ or $\omega_n^{(k)}=\omega_n-k(\tilde\Delta_1+
\tilde\Delta_2)$ repeating with a period 
$\tilde\Delta=\tilde\Delta_1+\tilde\Delta_2$. Thus, for a crude 
description
of the spectrum away from the Fermi edge it is convenient 
to divide the frequency region $\omega<0$ into the intervals
of width $\tilde\Delta$.

The number of peaks within the spectral interval 
$\{-\vert\omega\vert,-\vert\omega\vert-\tilde\Delta\}$ is the number of
integers satisfying the conditions 
$\vert\omega\vert < m\tilde\Delta_1+ n\tilde\Delta_2<|\omega|+\tilde\Delta$.
This number is equal to 
\begin{equation}
\label{number}
{\cal N}_{\omega}=
\frac{|\omega|\tilde\Delta}{2\tilde\Delta_1\tilde\Delta_2},
\end{equation}
where we assumed $\vert\omega\vert\gg \tilde\Delta$ and took into account the
parity restriction.
From Eq.\ (\ref{number}) we 
find the peak density 
$g_{\omega}=
{\cal N}_{\omega}/\tilde\Delta =\vert\omega\vert/2\tilde\Delta_1
\tilde\Delta_2$.
It also follows from ({\ref{number}) 
that ${\cal N}_{\omega-\tilde\Delta}-{\cal N}_{\omega}
=\tilde\Delta^2/2\tilde\Delta_1\tilde\Delta_2$
generations start within each interval. 
Since the heights of consecutive peaks within the
interval $\tilde\Delta$ vary non-monotonically, it is natural to
characterize these heights by the distribution function 
\begin{equation}
\label{dist}
F({\cal C})=
\frac{1}{2g_{\omega}}
\!
\int_0^{\infty}dm\, dn
\,\delta\bigl(\omega+m\tilde\Delta_1+n\tilde\Delta_2\bigr)\,
\delta\bigl(C_{mn}-{\cal C}\bigr),
\end{equation}
where $C_{mn}$ is given by Eq.\ (\ref{asym}). 
Here we made use of the fact that ${\cal N}_{\omega}\gg 1$ by treating
$m$ and $n$ as continuous variables. The prefactor in Eq.\ (\ref{dist})
ensures the normalization 
$\left(\int_0^{\infty}d{\cal C} F({\cal C})=1\right)$. It is easy to
see that $F({\cal C})$ is nonzero in the interval between 
${\cal C}_{min}=
{\rm min}\bigl\{2\mu_{1,2}
\bigl|\frac{\tilde\Delta_{1,2}}{\omega}\bigr|^{1-2\mu}\bigr\}$ 
and 
${\cal C}_{max}=
2\bigl|\frac{\omega}{\tilde\Delta}\bigr|^{\mu}{\rm max}\{\mu_{1,2}\}$
(we omit the overall factor $\epsilon^{2\mu}$).
Within this wide interval,
$F({\cal C})$ falls off as $\bigl({\cal C}_0/{\cal C}\bigr)^{2+\mu}$,
where 
\begin{equation}
{\cal C}_0=
\mu\biggl|\frac{\omega}{4}\bigl(\tilde{\Delta}_1^{-1}
+\tilde{\Delta}_2^{-1} \bigr)\biggr|^{2\mu-1}
\end{equation}
is the {\em typical} value of the oscillator strength. On the other hand,
the {\em average} oscillator strength, which can be easily calculated
from  Eq.\ (\ref{dist}), is equal to 
$\overline{\cal C}=\mu^{-1}{\cal C}_0\gg {\cal C}_0$.
The distribution function $F({\cal C})$ is schematically depicted in
Fig.\ \ref{fig:2}.  The fact that $\overline{\cal C}$ {\em decreases} with 
$|\omega|$ can be understood in the following way. As it is seen from
Eq.\ (\ref{cN}), in the symmetric case, with only a single
peak per interval $\tilde\Delta$, the peak heights increase with  
$|\omega|$ as $|\frac{\omega}{\tilde\Delta}|^{2\mu}$. 
In the general case,
this spectral intensity gets redistributed between ${\cal N}_{\omega}$
different peaks. Thus, 
\begin{equation}
\overline{\cal C} 
\sim {\cal N}_{\omega}^{-1}\biggl|\frac{\omega}{\tilde\Delta}\biggr|^{2\mu}
\propto |\omega|^{2\mu-1}.
\end{equation}
%

\section{Conclusions}

In the present paper we derived the emission spectrum from a highly
excited ring-shaped quantum dot. 
In this system electron-electron, hole-hole and 
electron-hole interactions relax the momentum  conservation
leading to a  multitude of discrete emission lines. 
Luttinger liquid model employed in  our calculation allows 
to evaluate the overlap integrals between the correlated initial 
and final many-body states. These overlap integrals determine the 
intensity of the corresponding spectral lines.

The theoretical value of the dimensionless interaction
parameter $\mu$ is determined by the ratio of
screened interaction $U$ to the level spacings 
$\tilde{\Delta}_1$ and $\tilde{\Delta}_2$ at the corresponding Fermi levels.
Both quantities depend on the number of excited carriers, $N$, which in
turn is determined by the excitation intensity. This, and
the sensitivity of the screening to the details of experimental
setup, lead to a common ambiguity in the theoretical determination of
$\mu$. For example, in quantum wires, the value of $\mu$ measured in resonant
tunneling experiments,\cite{yacoby00a,yacoby00b}
was significantly larger then theoretical estimates.
Concerning the estimates for $\tilde{\Delta}_1$ and $\tilde{\Delta}_2$,
in the experimental paper Ref.\ \onlinecite{warburton00}
on luminescence from ring-shape dots, the total energy separation
$\tilde{\Delta}$ between the lowest level was approximately $5$ meV.
This value comes almost exclusively from the conduction band,
due to the large ratio of the electron and hole effective masses.
Both $\tilde{\Delta}_1$ and $\tilde{\Delta}_2$ increase linearly with
increasing $N$. This implies that  the shake-up processes within
the hole system are experimentally much more relevant than those
for electrons.  
 
Note finally, that for emission from a finite electron-hole 1D system
considered here, the physics underlying the interaction-induced 
multiplication of the number of lines with departure from the Fermi
level is analogous to that for tunneling into a disordered quantum
dot.\cite{altshuler97}

\acknowledgements

Discussions with E. Ehrenfreund and D. Gershoni
are  gratefully acknowledged.
The work at Vanderbilt was supported by ONR Grant No. N00140010951.
The work in Utah was supported by NSF  Grant No. INT-0003710, the
Petroleum Research Fund  under  
ACS-PRF Grant No. 34302-AC6, and by the Army Research Office
under Grant No. DAAD 19-0010406. 


\appendix

\section{}

\label{app:a}

Here we outline the calculation of the  Green function (\ref{D-bos}) using a
bosonisation scheme for the multicomponent Luttinger liquid on a
ring.\cite{shahbazyan97} 
The right/left fermion fields are presented as
\begin{equation}
\label{fer}
\psi_{j\alpha}(x)
=(2\pi \epsilon)^{-1/2}e^{i\varphi_{j\alpha}(x)+i\alpha \pi x/L},
\end{equation}
where right/left ($\alpha=\pm$) bosonic fields $\varphi_{j\alpha}(x)$ are
related to the corresponding densities as 
$\rho_{j\alpha}(x)=\frac{\alpha}{2\pi}
\frac{\partial \varphi_{j\alpha}(x)}{\partial x}$
(here $\epsilon$ is a cutoff).
The bosonic field has a decomposition
\begin{equation}
\label{bos-free}
\varphi_{j\alpha}(x)=\varphi_{j\alpha}^0+\alpha N_{j\alpha}2\pi x/L
+\bar{\varphi}_{j\alpha}(x),
\end{equation}
where the number operator $N_{j\alpha}$ and its conjugate $\varphi_{j\alpha}^0$
satisfy the commutation relations
\begin{equation}
\label{zero-comm}
[N_{j\alpha},\varphi_{l\beta}^0]=i\delta_{jl}\delta_{\alpha\beta}, 
\end{equation}
and the periodic fields
$\bar{\varphi}_{j\alpha}(x)=\bar{\varphi}_{j\alpha}(x+L)$ have the usual form, 
\begin{equation}
\label{bos-free-per}
\bar{\varphi}_{j\alpha}(x)=
\sum_{q}\theta(q\alpha)\sqrt{\frac{2\pi}{L|q|}}e^{-|q|\epsilon/2}
\Bigl( e^{iqx}a_{qj} + e^{-iqx}a_{qj}^{\dag}\Bigr),
\end{equation}
with $a_{qj}$ and $a_{qj}^{\dag}$ satisfying standard boson commutation
relations [$\theta(x)$ is the step function]. The boundary condition for the
fermion fields, $\psi_{j\alpha}(x+L)=(-1)^{N_j}\psi_{j\alpha}(x)$,
depends on the parity of the number of particles, $N_j=2N_{j\alpha}$.
The Hamiltonian $H=H_0+H_{int}$ is quadratic in boson fields:
\begin{equation}
\label{H-free}
H_0=\sum_{j\alpha}\frac{v_j}{4\pi}
\int_0^Ldx
\Biggl[\frac{\partial \varphi_{j\alpha}(x)}{\partial x}\Biggr]^2,
\end{equation}
and 
\begin{eqnarray}
\label{H-int}
H_{int}=\frac{1}{2}\sum_{jl}
\int_0^Ldx\int_0^Ldy
&&
\Biggl[\sum_{\alpha}\frac{\alpha}{2\pi}
\frac{\partial \varphi_{j\alpha}(x)}{\partial x}\Biggr]
U_{jl}(x-y)
\nonumber\\
&&
\times
\Biggl[\sum_{\beta}\frac{\beta}{2\pi}
\frac{\partial \varphi_{l\beta}(y)}{\partial y}\Biggr],
\end{eqnarray}
where $U_{jl}(x)$ is the screened potential. Using 
Eqs.\ (\ref{bos-free}) and (\ref{bos-free-per}), and after separating out the
zero-mode part of the Hamiltonian, $H^0$, from the bosonic part, $\bar{H}$, the
total Hamiltonian $H=H^0+\bar{H}$ can be written as
\begin{eqnarray}
\label{H-total}
H=
&&
\frac{\pi}{L}\sum_{jl\alpha\beta}
N_{j\alpha}
\biggl(v_j\delta_{jl}\delta_{\alpha\beta}+\frac{u_{jl}}{2}\biggl)
N_{l\beta}
\nonumber\\
&&
+
\sum_{qjl} e^{-|q|\epsilon}|q|
\biggl[v_j\delta_{jl}a_{qj}^{\dag}a_{qj}
\nonumber\\
&&
\mbox{\hspace{20mm}}
+\frac{u_{jl}}{4}(a_{qj}^{\dag}+a_{-qj})(a_{-ql}^{\dag}+a_{ql})
\biggr],
\end{eqnarray}
where $u_{jl}=\pi^{-1}\int dx U_{jl}(x)$.

In order to calculate correlation functions, the Hamiltonian $\bar{H}$, 
corresponding to the second term of (\ref{H-total}), has to be brought to
the canonical form. This is done in two steps. First, we perform a
two-component Bogolubov's transformation in order to 
eliminate the cross-terms with opposite momenta,
\begin{eqnarray}
\label{bogol}
&&
a_{qj}=\sum_l\bigl(X_{jl}b_{ql}+Y_{jl}b_{-ql}^{\dag}\bigr),
\nonumber\\
&&
\sum_l\bigl(X_{jl}X_{ln}^{\dag}-Y_{jl}Y_{ln}^{\dag}\bigr)=\delta_{jn}.
\end{eqnarray}
We then obtain
\begin{eqnarray}
\label{H-bar1}
\bar{H}=\sum_{qjl} e^{-|q|\epsilon}|q|
b_{qj}^{\dag}\bigl(X^{\dag}-Y^{\dag}\bigl)_{jl}v_l
\bigl(X-Y\bigl)_{ln}b_{qn},
\end{eqnarray}
where the matrices $X$ and $Y$ must satisfy
\begin{eqnarray}
\label{bogol-eq}
\sum_{lm}
\bigl(X^{\dag}+Y^{\dag}\bigl)_{jl}
&&
(u_{lm}+v_l\delta_{lm})
\bigl(X+Y\bigl)_{mn}
\nonumber\\
&&
=
\sum_l
\bigl(X^{\dag}-Y^{\dag}\bigl)_{jl}v_l
\bigl(X-Y\bigl)_{ln}.
\end{eqnarray}
Second, we diagonalize the Hamiltonian (\ref{H-bar1}) by first presenting the
matrices $X$ and $Y$ as 
\begin{eqnarray}
\label{XY}
X=\cosh\lambda \,O,
~~~
Y=\sinh\lambda\, O,
\end{eqnarray}
where $\lambda_{jl}=\lambda_j\delta_{jl}$ is diagonal matrix of Bogolubov's 
angles $\lambda_j$ and $O$ is an orthogonal matrix, and then by introducing new
boson operators $c_{qj}=\sum_lO_{jl}b_{ql}$. The Hamiltonian $\bar{H}$ then
takes the form
\begin{eqnarray}
\label{H-diag}
\bar{H}=\sum_{qj} e^{-|q|\epsilon}|q| \tilde{v}_jc_{qj}^{\dag}c_{qj},
\end{eqnarray}
with renormalized Fermi velocities $\tilde{v}_j=e^{-2\lambda_j}v_j$. The
old and new boson operators are related as
\begin{eqnarray}
\label{old-new}
a_{qj}=\cosh\lambda_j\, c_{qj}+ \sinh\lambda_j\, c_{-qj}^{\dag}.
\end{eqnarray}
Using the decomposition (\ref{XY}), Eq.\ (\ref{bogol-eq}) takes the
form $\tilde{O}AO=0$, where $\tilde{O}$ is the transposed matrix,
and the matrix $A$ is given by
\begin{eqnarray}
\label{bogol-matr}
A_{jl}=u_{jl}e^{\lambda_j+\lambda_l}+
\delta_{jl}v_j\bigl(e^{2\lambda_j}-e^{-2\lambda_j}\bigr).
\end{eqnarray}
The Bogolubov's angles $\lambda_j$ are found from the condition
that all the eigenvalues of $A_{jl}$ vanish.
In the two-component case, this yields
\begin{eqnarray}
\label{two-comp}
&&
e^{-2\lambda_1}=
\sqrt{Q
\frac{v_1+u_{11}-v_2 Q}
{v_1 Q-v_2-u_{22}}},
\mbox{\hspace{5mm}}
e^{-2\lambda_2}=Q /e^{-2\lambda_1},
\nonumber\\ &&
Q=\sqrt{\biggl(1+\frac{u_{11}}{v_1}\biggr) 
\biggl(1+\frac{u_{22}}{v_2}\biggr)
-\frac{u_{12}^2}{v_1v_2}}.
\end{eqnarray}
The Luttinger liquid interaction parameter is given by
$\mu_j=\sinh^2\lambda_j$. 
In the case of weak interactions,
$u_{jl}/v_j\ll 1$, we have $\lambda_j\simeq -u_{jj}/4v_j$ so that
$\mu_j\simeq \lambda_j^2 \simeq (u_{jj}/4 v_j)^2$ and 
$\tilde{\Delta}_j\simeq \Delta_j(1+u_{jj}/2v_j)$.

With the Hamiltonian (\ref{H-diag}), the time-dependence of new operators is
standard, $c_{qj}(t)=e^{-i\tilde{v}_j|q|t}c_{qj}$. 
Using the relation (\ref{old-new}), the periodic fields (\ref{bos-free-per})
take the form 
\begin{eqnarray}
\label{bos-int-per}
\bar{\varphi}_{j\alpha}(x,t)=
&&
\sum_{q}
\sqrt{\frac{2\pi}{L|q|}}e^{-|q|\epsilon/2}
\nonumber\\
&&
\times
\Bigl[\theta(q\alpha)\cosh\lambda_j
+ \theta(-q\alpha)\sinh\lambda_j\Bigr]
\nonumber\\
&&
\times
\Bigl(e^{iqx-i\tilde{v}_j|q|t}c_{qj}
+e^{-iqx+i\tilde{v}_j|q|t}c_{qj}^{\dag}\Bigr).
\end{eqnarray}
The time-dependence of zero-modes is governed by 
the zero-mode part (first term) of the Hamiltonian
(\ref{H-total}). The time-dependent bosonic field is finally obtained as
\begin{eqnarray}
\label{bos-int}
\varphi_{j\alpha}(x,t)=
&&
\varphi_{j\alpha}^0
+\alpha N_{j\alpha}2\pi (x-\alpha v_j t)/L
\nonumber\\
&&
-\sum_{l\beta}u_{jl}N_{l\beta}\,\pi t/L
+\bar{\varphi}_{j\alpha}(x,t).
\end{eqnarray}
We are now in position to calculate the Green functions. For this,
we separate out annihilation and creation parts of the periodic field
(\ref{bos-int-per}), 
$\bar{\varphi}_{j\alpha}(x,t)=\bar{\varphi}_{j\alpha}^{-}(x,t)
+\bar{\varphi}_{j\alpha}^{+}(x,t)$, which satisfy the following commutation
relations 
\begin{eqnarray}
\label{bos-int-per-comm}
[\bar{\varphi}_{j\alpha}^{-}(x,t),\bar{\varphi}_{j\alpha}^{+}(x',t')]
=
&&
\ln f_{\alpha}(z_{j\alpha}-z'_{j\alpha})
\nonumber\\
&&
\mbox{\hspace{-30mm}}
+\mu_j\ln \Bigl[
f_{\alpha}(z_{j\alpha}-z'_{j\alpha}) f_{-\alpha}(z_{j,-\alpha}-z'_{j,-\alpha})
\Bigr],
\end{eqnarray}
with $z_{j\alpha}=x-\alpha\tilde{v}_jt$. Then we present the fermion operator
(\ref{fer}) in the normal-ordered form,
\begin{eqnarray}
\label{fer-normal}
\psi_{j\alpha}(x,t)=
&&
\psi_{j\alpha}^{0}(x,t)\bar{\psi}_{j\alpha}(x,t),
\nonumber\\
\psi_{j\alpha}^{0}(x,t)=
&&
e^{iv_j(1+u_{jj}/2)\pi t/L}e^{i\varphi_{j\alpha}^0}
\nonumber\\
&&
\times
e^{i\alpha N_{j\alpha}2\pi z_{j\alpha}/L
-i\sum_{l\beta}u_{jl}N_{l\beta}\,\pi t/L},
\nonumber\\
\bar{\psi}_{j\alpha}(x,t)=
&&
L^{-1/2}\Bigl(2\pi\epsilon/L\Bigr)^{\mu_j}
e^{i\bar{\varphi}_{j\alpha}^{+}(x,t)}
e^{i\bar{\varphi}_{j\alpha}^{-}(x,t)},
\end{eqnarray}
where we again separated out zero-mode and periodic parts.
Using Eq.\ (\ref{fer-normal}) together with commutators 
(\ref{zero-comm}) and (\ref{bos-int-per-comm}), 
the Green function (\ref{prob-D}) can be straightforwardly calculated as
\begin{eqnarray}
\label{D-bos-app}
D_{\alpha}(x,t)=
&&
\biggl(\frac{2\pi\epsilon}{L}\biggr)^{2(\mu_2+\mu_2)}
\frac{e^{-it\delta_P-it\delta_u}}{L^2}
\nonumber\\ && \times
\Bigl[f_{\alpha}(z_{1\alpha})\Bigr]^{1+\mu_1}
\Bigl[f_{-\alpha}(z_{1,-\alpha})\Bigr]^{\mu_1}
\nonumber\\ && \times
\Bigl[f_{-\alpha}(z_{2,-\alpha})\Bigr]^{1+\mu_2}
\Bigl[f_{\alpha}(z_{2\alpha})\Bigr]^{\mu_2},
\end{eqnarray}
where $\delta_P=\pi (v_1+v_2)/L$ and $\delta_u=\pi (u_{11}+u_{22}+2u_{12})/2$
are the energy shifts due the changes in the
parity of electron and hole numbers and in the Coulomb energy,
caused by a removal of an {\em e-h} pair. We assume that the screened
interaction is the same for electrons and holes, $u_{11}=u_{22}=-u_{12}=u$, so
that $\delta_u=0$. Then, after absorbing the factor $2\pi/L$ into $\epsilon$,
we arrive at Eq.\ (\ref{D-bos}). 

Note finally that the above calculation is easily generalized if the ring is
penetrated by a magnetic flux $\phi$. In this case, the electron and hole
number operators should be shifted by flux-dependent constants,
$N_{1\alpha}\rightarrow N_{1\alpha}+\alpha\phi/\phi_0$ and
$N_{2\alpha}\rightarrow N_{2\alpha}-\alpha\phi/\phi_0$,
where $\phi_0$ is the flux quantum. This results in a replacement 
$\delta_P\rightarrow \delta_P(1-2\alpha\phi/\phi_0)$ in Eq.\ (\ref{D-bos}).

\section{}

\label{app:b}

Substituting the Fourier expansion
\begin{eqnarray}
\label{I-fourier}
\Bigl[f_{\alpha}(z_{j\alpha})\Bigr]^{\nu}=
&&
\sum_n
b_{\nu}(n)e^{i\alpha 2\pi nz_{j\alpha}/L},
\nonumber\\
b_{\nu}(n)=
&&
\frac{\sin \pi\nu}{\pi}
B(n+{\nu},1-{\nu}),
\end{eqnarray}
into Eq.\ (\ref{D-bos}), $D_{\alpha}(\omega)$ takes the form
\begin{eqnarray}
\label{D-moment}
D_{\pm}(\omega)=
\epsilon^{2(\mu_1+\mu_2)}
\sum_{\{n\}}
&&
b_{1+\mu_1}(n_1)b_{\mu_1}(n'_1)b_{1+\mu_2}(n_2)
\nonumber\\
&&\times
b_{\mu_2}(n'_2)
\Lambda_{\pm}(\omega,\{n\}),
\end{eqnarray}
with
\begin{eqnarray}
\label{Lambda}
\Lambda_{\pm}(\omega,\{n\})
&&
=\frac{1}{L^2}\int dt \int_0^{L} dx
\exp\biggl[
-i\omega t
\nonumber\\ &&
 \pm \, i\frac{2\pi}{L}
\Bigl(n_1z_{1\pm} -n'_1z_{1\mp}
-n_2z_{2\mp}+n'_2z_{2\pm}
\Bigr)
\biggr]
\nonumber\\ &&
=\frac{2\pi}{L}
\delta_{n_1-n'_1,n_2-n'_2}
\delta\biggl[\omega+\frac{2\pi \tilde{v}_1}{L}\Bigl(n_1+n'_1\Bigr)
\nonumber\\ &&
\mbox{\hspace{26mm}}
+\frac{2\pi \tilde{v}_2}{L}\Bigl(n_2+n'_2\Bigr)
\biggr],
\end{eqnarray}
where we absorbed the parity shift $\delta_P$ into $\omega$.
The Kroniker delta and the delta-function reflect the conservation of 
momentum and energy, respectively. Thus, we obtain
\begin{eqnarray}
\label{D-final}
D_{\alpha}(\omega)=\frac{2\pi}{L}\sum_{mn}C_{mn}
\delta\Bigl(\omega +\tilde{\Delta}_1m+\tilde{\Delta}_2n\Bigr),
\end{eqnarray}
with
\begin{eqnarray}
\label{Cmn}
C_{mn}=
\epsilon^{2(\mu_1+\mu_2)}
\sum_{l}
&&
b_{1+\mu_1}[(m+n)/2-l]b_{1+\mu_2}(n-l)
\nonumber\\ && \times
b_{\mu_1}[(m-n)/2+l]b_{\mu_2}(l).
\end{eqnarray}
Finally, using the integral representation for the Beta-function in 
Eq.\ (\ref{I-fourier}) we arrive at Eq.\ (\ref{Cmn-int}).
The sum in Eq.\ (\ref{D-final}) is constrained by the selection rule
that $m$ and $n$ are of the same parity, as can be seen from  
Eq.\ (\ref{Lambda}). From Eq.\ (\ref{D-final}), the emission
spectrum (\ref{modified}) follows.


\section{}

\label{app:c}


Here we consider the case when
the level spacings in the conduction and valence bands are commensurate:
$\tilde{\Delta}_1/\tilde{\Delta}_2=p/q$, where $p$ and $q$ are integers.
Then Eq.\ (\ref{D-final}) takes the form
\begin{eqnarray}
\label{D-rat-gen}
D_{\alpha}(\omega)
&&
=\frac{2\pi}{L}\sum_{mn}C_{mn}
\delta\biggl(\omega +\tilde{\Delta}\frac{mp+nq}{p+q}\biggr)
\nonumber\\
&&
=\frac{2\pi}{L}\sum_k C_k\delta(\omega+\tilde{\Delta} k/Q),
\end{eqnarray}
where $Q=p+q$, $P=p-q$, $\tilde{\Delta}=\tilde{\Delta}_1+\tilde{\Delta}_2$,
and
\begin{eqnarray}
\label{Ck-gen}
C_k
&&
=\sum_{mn}\delta_{k,mp+nq}C_{mn}
=\sum_{MN}\delta_{k,MP+NQ}C_{N+M,N-M}
\nonumber\\
&&
=\sum_{MN}\delta_{k-MP,NQ}
C_{\frac{k}{Q}+M\bigl(1-\frac{P}{Q}\bigr),
\frac{k}{Q}-M\bigl(1+\frac{P}{Q}\bigr)}.
\end{eqnarray}
Using the relation
\begin{eqnarray}
\sum_N\delta_{k,NQ}=\frac{1}{Q}\sum_{l=0}^{Q-1}e^{-i2\pi l k/Q},
\end{eqnarray}
the oscillator strengths
can be presented as
\begin{eqnarray}
\label{Ck-final-gen}
C_k=\frac{1}{Q}\sum_{l=0}^{Q-1} e^{-i2\pi l k/Q}f_l(k),
\end{eqnarray}
with
\begin{eqnarray}
\label{f-genl}
f_l(k)=\sum_Me^{i2\pi l M P/Q}
C_{\frac{k}{Q}+M\bigl(1-\frac{P}{Q}\bigr),
\frac{k}{Q}-M\bigl(1+\frac{P}{Q}\bigr)}.
\end{eqnarray}
Using integral representation (\ref{Cmn-int}), the sum over $M$ can be
explicitly performed. For $k/Q=|\omega|/\tilde{\Delta}\gg 1$, the
resulting expression for coefficients $f_l$ takes the form
\begin{eqnarray}
\label{fl-int-asympt-gen}
f_l(k)
=
&&
\int_{-\infty}^{\infty}\frac{d\phi_1d\phi_2}{(2\pi)^2}
\frac{\epsilon^{2(\mu_1+\mu_2)}\, e^{-i(\phi_1+\phi_2)k/Q}}
{(-i\phi_1)^{1+\mu_1}(-i\phi_2)^{1+\mu_2}}
\nonumber\\&&\times
\frac{1}{\Bigl(1-s_l-is_l[\phi_2+(\phi_1+\phi_2)P/Q]\Bigr)^{\mu_1}}
\nonumber\\&&\times
\frac{1}
{\Bigl(1-s_l^{\ast}-is_l^{\ast}[\phi_1-(\phi_1+\phi_2)P/Q]\Bigr)^{\mu_2}},
\end{eqnarray}
where $s_l=e^{i2\pi lP/Q}$. The $l$-dependence of $f_l(k)$ is 
determined by the relative magnitude of $Q/k$ and $|1-s_l|$:
\begin{eqnarray}
\label{fl-est1}
f_l(k)
\simeq 
\biggl|\frac{\epsilon k}{Q}\biggr|^{2(\mu_1+\mu_2)}
\end{eqnarray}
for $k/Q\ll |1-s_l|^{-1}$, and
\begin{eqnarray}
\label{fl-est2}
f_l(k)\simeq
\Biggl(\frac{\epsilon^2 k/Q}{1-s_l}\Biggr)^{\mu_1}
\Biggl(\frac{\epsilon^2 k/Q}{1-s_l^{\ast}}\Biggr)^{\mu_2}
\end{eqnarray}
for  $k/Q\gg |1-s_l|^{-1}$,
with the two estimates matching at $k/Q\sim |1-s_l|^{-1}$.

In the case $\tilde{\Delta}_1/\tilde{\Delta}_2=3$,
corresponding to $P=2$ and $Q=4$ so that $s_l=(-1)^l$, the coefficients $f_l$
take two different values depending on the parity of $l$, 
\begin{eqnarray}
\label{fl-even}
&&
f_{even}(k)
\simeq 
\biggl|\frac{\epsilon k}{4}\biggr|^{2(\mu_1+\mu_2)}
=\biggl|\frac{\epsilon \omega}{\tilde{\Delta}}\biggr|^{2(\mu_1+\mu_2)},
\\
\label{fl-odd}
&&
f_{odd}(k)\simeq
\biggl|\frac{\epsilon^2 k}{8}\biggr|^{\mu_1+\mu_2}
=\biggl|\frac{\epsilon^2 \omega}{2\tilde{\Delta}}\biggr|^{\mu_1+\mu_2},
\end{eqnarray}
yielding
\begin{eqnarray}
\label{Ck-even-odd}
C_k=
\biggl|\frac{\epsilon \omega}{\tilde{\Delta}}\biggr|^{2\mu}
\frac{1+(-1)^k}{2}
\frac{1+e^{i\pi k/2}\bigl|\frac{2\omega}{\tilde{\Delta}}\bigr|^{-\mu}}{2}
\end{eqnarray}
with $\mu=\mu_1+\mu_2$. Obviously, $C_k=0$ for $k$ odd. For $k$ even, we have
\begin{eqnarray}
\label{Ck-even-even}
&&
C_{4l}\simeq
\biggl|\frac{\epsilon \omega}{\tilde{\Delta}}\biggr|^{2\mu}
\frac{1+\bigl|\frac{2\omega}{\tilde{\Delta}}\bigr|^{-\mu}}{2},
\\
&&
C_{4l+2}\simeq
\biggl|\frac{\epsilon \omega}{\tilde{\Delta}}\biggr|^{2\mu}
\frac{1-\bigl|\frac{2\omega}{\tilde{\Delta}}\bigr|^{-\mu}}{2},
\end{eqnarray}
leading to Eq.\ (\ref{D=2}).


$^*$On leave from Vanderbilt University


\begin{figure}
\caption{Emission spectrum for $\tilde\Delta_1=3\tilde\Delta_2$. 
In the low-frequency domain, the single-particle peaks acquire weak 
many-body satellites; in the high-frequency domain, the heights of 
the single-particle ($|\omega|/\tilde\Delta=l$) and many-body
($|\omega|/\tilde\Delta=l+1/2$) peaks are close to each other. 
Inset: Single-particle spectra of electrons and hole in conduction 
and valence bands, respectively; $\omega_{th}$ is the energy 
distance between the corresponding Fermi levels.
}
\label{fig:1}
\end{figure}

\begin{figure}
\caption{
The distribution function Eq.\ (\ref{dist}) of the peak heights within the 
interval $\tilde\Delta$ is plotted schematically versus 
$x={\cal C}/{\cal C}_0$. The minimal value of $x$ is $x_{min}\sim 1$, while 
$x_{max}\sim \bigl|\omega/\tilde\Delta\bigr|^{1-\mu}\gg 1$. The point
$x=\mu^{-1}$ corresponds to the average oscillator strength.
}
\label{fig:2}
\end{figure}

\clearpage

\begin{center}
\epsfxsize=6.0in
\epsffile{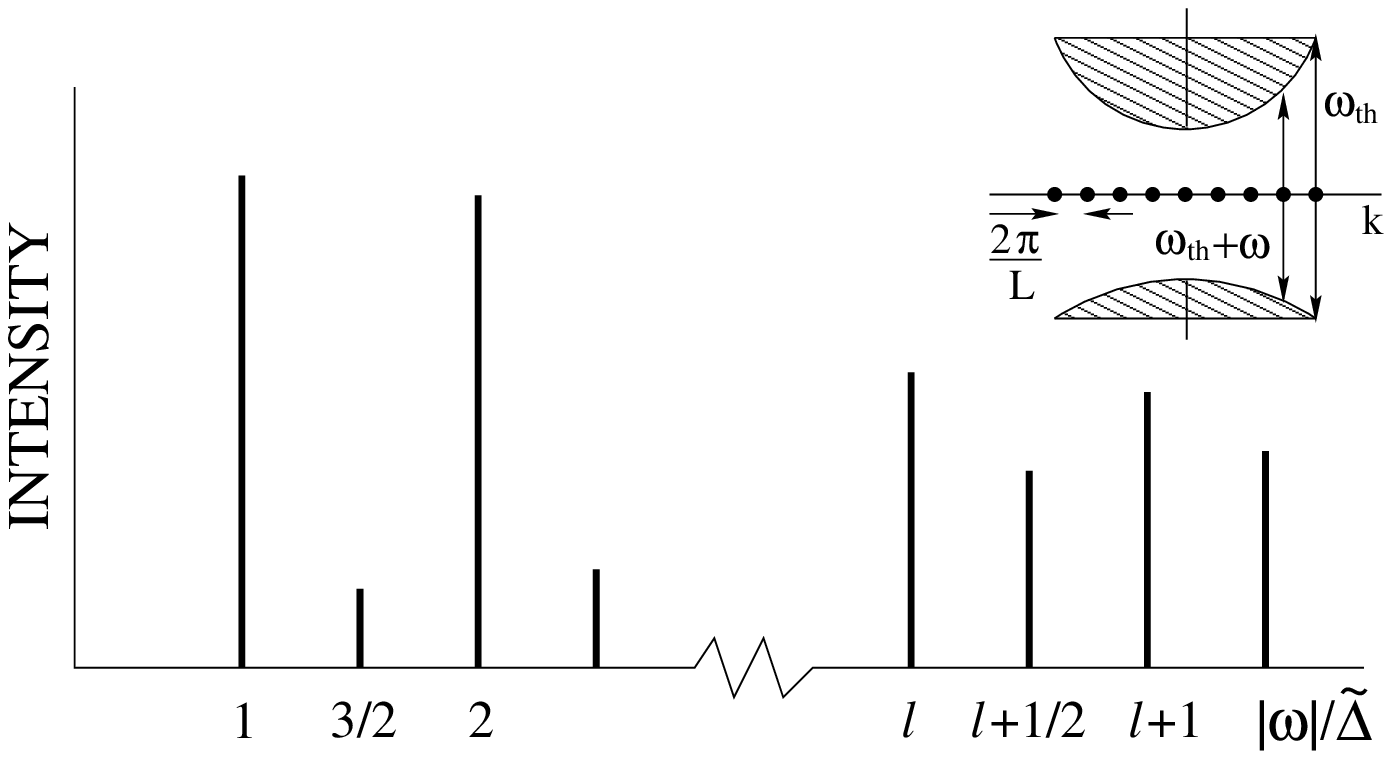}
\end{center}
\vspace{80mm}
\centerline{Fig. 1}

\clearpage

\begin{center}
\epsfxsize=6.0in
\epsffile{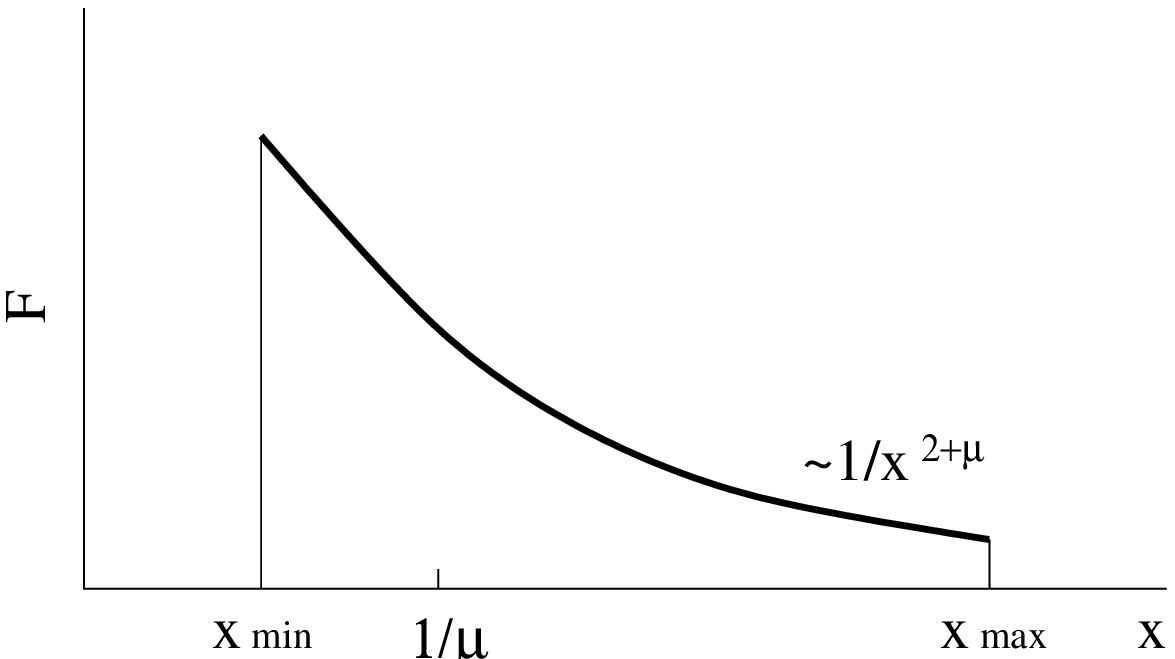}
\end{center}
\vspace{80mm}
\centerline{Fig. 2}


\begin{references}
\bibitem{brunner92}
K. Brunner,
U. Bockelmann, G. Abstreiter, M. Walther, G. Bohm, G. Trankle, and G. Weimann
Phys.\ Rev.\ Lett.\ {\bf 69},\ 3216\ (1992).
%
\bibitem{marzin94}
J.-Y. Marzin, J.-M. G\'{e}rard, A. Izra\"{e}l,
D. Barrier, and G. Bastard, 
Phys.\ Rev.\ Lett.\ {\bf 73},\ 716\ (1994).
%
\bibitem{zrenner00}A. Zrenner, J. Chem. Phys.
{\bf 112}, 7790 (2000).

\bibitem{gammon00}
D. Gammon,
Nature\ {\bf 405}\, 899\ (2000).

\bibitem{warburton00}
R. J. Warburton,
C. Sch\"{a}flein, D. Haft, F. Bicken, A. Lorke,
K. Karrai, J. M. Garcia, W. Schoenfeld, and P. M. Petroff,
Nature\ {\bf 405},\ 926\ (2000).

\bibitem{regelman01} D. V. Regelman, E. Dekel,
D. Gershoni, E. Ehrenfreund, A. J. Williamson, J. Shamway, A. Zunger,
W. V. Schoenfeld, and P. M. Petroff, ArXiv: cond-mat 0105589.

\bibitem{ikezawa97}
M. Ikezawa,
Y. Masumoto, T. Takagahara, and S. V. Nair,
Phys.\ Rev.\ Lett.\ {\bf 79},\ 3522\ (1997).
%
\bibitem{landin98}
L. Landin
M. S. Miller, M.-E. Pistol, C. E. Pryor, and L. Samuelson,
Science\ {\bf 280},\ 262\ (1998).
%
\bibitem{bayer98}
M. Bayer,
T. Gutbrod, A. Forchel, V. D. Kulakovskii,
A. Gorbunov,  M. Michel, R. Steffen, and K. H. Wang,
Phys.\ Rev.\ B\ 58,\ 4740\ (1998).
%
\bibitem{kulakovskii99}
V. D. Kulakovskii,
G. Bacher, R. Weigand,
T. Kummell, A. Forchel, E. Borovitskaya, K. Leonardi, and D. Hommel, 
Phys.\ Rev.\ Lett.\ {\bf 82},\ 1780\ (1999).

\bibitem{dekel98}
E. Dekel,
D. Gershoni, E. Ehrenfreund, D. Spektor,
J. M. Garcia, and P. M. Petroff, 
Phys.\ Rev.\ Lett.\ {\bf 80},\ 4991\ (1998). 
%
\bibitem{dekel00a}
E. Dekel, D. Gershoni, E. Ehrenfreund, J. M. Garcia, and P. M. Petroff, 
Phys.\ Rev.\ B {\bf 61},\ 11009\ (2000).
%
\bibitem{dekel00b}
E. Dekel, D. Regelman, D. Gershoni, E. Ehrenfreund, 
W. V. Schoenfeld, and P. M. Petroff, 
Phys.\ Rev.\ B {\bf 62},\ 11038\ (2000).
%
\bibitem{dekel01}
E. Dekel, D. Regelman, D. Gershoni, E. Ehrenfreund, 
W. V. Schoenfeld, and P. M. Petroff,
Solid State Commun.\ {\bf 117}, 395 (2001).
%
\bibitem{findeis00}F. Findeis, A. Zrenner, G. Bohm, and G. Abstreiter,
Solid\ State\ Commun.\ {\bf 114},\ 227\ (2000).

\bibitem{bayer00}
M. Bayer,
O. Stern, P. Hawrylak, S. Fafard, and A. Forchel, 
Nature\ {\bf 405},\ 923\ (2000).

\bibitem{hawrylak99}
P. Hawrylak, 
Phys.\ Rev\. B\ {\bf 60},\ 5597\ (1999).

\bibitem{Toda}
Y. Toda, O. Moriwaki, M. Nishioka, and Y. Arakawa,
Phys.\ Rev.\ Lett.\ {\bf 82},\ 4114\ (1999).

\bibitem{lorke00}
A. Lorke,
R. J. Luyken, A. O. Govorov,
J. P. Kotthaus, J. M. Garcia, and P. M. Petroff, 
Phys.\ Phys.\ Lett.\ {\bf 84},\ 2223\ (2000).
%
\bibitem{peterson00}
H. Pettersson
R. J. Warburton, A. Lorke, K. Karrai, J. P. Kotthaus,
J. M. Garcia, and P. M. Petroff, 
Physica\ E\ {\bf 6},\ 510\ (2000).
%
\bibitem{chaplik95}
A. Chaplik, 
Pis'ma\ Zh.\ Eksp.\ Teor.\ Fiz.\ {\bf 62},\ 885\ (1995) 
[JETP\ Lett.\ {\bf 62},\ 900\ (1995)].

\bibitem{rudo00}
R. A. R\"{o}mer and M. E. Raikh, 
Phys.\ Rev.\ B\ {\bf 62},\ 7045\ (2000).

\bibitem{HuiHu00}
H. Hu,
D.-J. Li, J.-L. Zhu, and J.-J. Xiong,
J.\ Phys.\ Condens.\ Matter\ {\bf 12},\ 9145\ (2000).
%
\bibitem{HuiHu01a}
H. Hu, G.-M. Zhang, J.-L. Zhu, and J.-J. Xiong, 
Phys.\ Rev.\ B\ {\bf 63},\ 045320\ (2001).
%
\bibitem{HuiHu01b}
H. Hu, J.-L. Zhu, D.-J. Li, and J.-J. Xiong, 
Phys.\ Rev.\ B\ {\bf 63},\ 195307\ (2001).

\bibitem{ulloa00}
J. Song and S. E. Ulloa,
Phys.\ Rev.\ B {\bf 63},\ 125302\ (2001).

\bibitem{schulz95}
See, e.g., H. J. Schulz, in 
{\em Proceedings of Les Houches Summer School LXI}, 
edited by E. Akkermans,
G. Montambaux, J. Pichard, and J. Zinn-Justin,
(Elsevier, Amsterdam, 1995), p. 533. 
%
\bibitem{gogolin93}
A.\ O.\ Gogolin,
Phys.\ Rev.\ Lett. {\bf 71},\ 2995\ (1993).
%
\bibitem{prokof'ev94}
N. V. Prokof'ev,
Phys.\ Rev.\ B {\bf 49},\ 2148\ (1994).
%
\bibitem{kane94}
C. L. Kane, K. A. Matveev, and L. I. Glazman,
Phys.\ Rev.\ B {\bf 49},\ 2253\ (1994).
%
\bibitem{sassetti98}
M. Sassetti and B. Kramer,
Phys.\ Phys.\ Lett.\ {\bf 80},\ 1485\ (1998).
%
\bibitem{kramer00}
B. Kramer and M. Sassetti,
Phys.\ Phys.\ B\ {\bf 62},\ 4238\ (2000).
%
\bibitem{larkin74}
I. E. Dzyaloshinsky and A. I. Larkin,
Zh.\ Eksp.\ Teor.\ Fiz. {\bf 65},\ 411\ (1974)
[Sov.\ Phys.\ JETP {\bf 38},\ 202\ (1974)].
%
\bibitem{matveev93}
K.\ A.\ Matveev and L.\ I.\ Glazman,
Phys.\ Rev.\ Lett. {\bf 70},\ 990\ (1993).
%
\bibitem{penc93}
K. Penc and J. S\'{o}lyom,
Phys.\ Rev.\ B {\bf 47},\ 6273\ (1993).

\bibitem{shahbazyan97}T.\ V.\ Shahbazyan and S.\ E.\ Ulloa, 
Phys.\ Rev.\ B {\bf 55},\ 13702\ (1997).

\bibitem{yacoby00a}
O.\ M.\ Auslaender,
A. Yacoby, R. de Picciotto, K. W. Baldwin, L. N. Pfeiffer, K. W. West,
Phys.\ Rev.\ Lett.\ {\bf 84},\ 1764\ (2000).

\bibitem{yacoby00b}
T. Kleimann,
M. Sassetti, B. Kramer, and A. Yacoby,
Phys.\ Rev.\ B\ {\bf 62},\ 8144\ (2000).

\bibitem{altshuler97}B.\ L.\ Altshuler,
Y.\ Gefen, A.\ Kamenev, and L.\ S.\ Levitov,
Phys.\ Rev.\ Lett.\ {\bf 78},\ 2803\ (1997). 
\end{references}
\end{document}